\documentclass[aps,prl,floatfix,twocolumn,showpacs]{revtex4} 

\usepackage{graphicx}

\begin{document} 

\title{Epidemic dynamics on an adaptive network} 

\author{Thilo~Gross} 
\author{Carlos~J.~Dommar~D'Lima}
\author{Bernd~Blasius}
\affiliation{AG Nichtlineare Dynamik, Institut f\"ur Physik, Universit\"at Potsdam,  Am Neuen Palais 10, 14469 Potsdam, Germany}
\date{\today} 

\begin{abstract} Many real world networks are characterized by adaptive
changes in their topology depending on the state of their nodes. Here we study
epidemic dynamics on an adaptive network, where the susceptibles are able to
avoid contact with infected  by rewiring their network connections. This gives
rise to assortative degree correlation, oscillations, hysteresis and 1st
order transitions. We propose a low-dimensional model to describe the system and
present a full local bifurcation analysis. Our results indicate that the
interplay between dynamics and topology can have important consequences for the
spreading of infectious diseases and related applications.  \end{abstract} 

\maketitle

In the physical literature the dynamics of complex networks has recently
received much attention, with many applications in social, biological and
technical systems  \cite{Barabasi:NetworkReview,Newman:Review}. In particular,
most research has been directed in two distinct directions.  On the one hand,
attention has been paid to the structure of the networks, revealing that simple
dynamical rules, such as preferential  attachment or selective rewiring, can be
used to generate complex topologies
\cite{Price:PowerLaws,Watts:SmallWorld,Barabasi:PowerLaw,Dorogovtsev:Growth}. Many of these rules 
are not only a useful tool for the generation of model graphs, but are also
believed to shape real-world  networks like the internet or the network of
social contacts.  On the other hand research has focused
on large ensembles of dynamical systems, where the interaction between
individual units is described by a complex graph 
\cite{Pecora:SmallWorldSync,Kuperman:SmallWorldEpidemics,PastorSatorras:ScaleFree,May:ScaleFree,
PastorSatorras:NoThresholds,Eguiluz:ScaleFree,Newman:Assortative,Newman:Mixing,Sokolov:Assortative}. 
These studies have shown that the network topology can have a strong impact on the dynamics 
of the nodes, e.g., the absence of epidemic thresholds on scale free networks
\cite{PastorSatorras:ScaleFree,May:ScaleFree}
or the detrimental effect of assortative degree correlations on targeted
vaccination \cite{Newman:Assortative}. In
the past the cross fertilization between these two lines of  thought has led to
considerable advances. However, \emph{the dynamics of  networks} and \emph{the
dynamics on networks} are still generally studied  separately. In doing so, a
characteristic features  of many real world networks is not taken into account,
namely the ability to adapt the network topology dynamically  in response to the
dynamic state of nodes
\cite{Bornholdt:AdaptiveNeurons,Bornholdt:AdaptiveNetworks,Holme:NetworkAgents,Zhou:AdaptiveNets}.

Consider for example the spreading of an infectious disease on a social network. Humans tend to respond to the emergence of an epidemic by avoiding contacts with infected individuals. Such rewiring of the local connections can have a strong effect on the dynamics of the disease, which in turn influences the rewiring process. Thus, a complicated mutual interaction between a time varying network topology and the dynamics of the nodes emerges. 

In this Letter we study a susceptible-infected-susceptible (SIS) model on an 
adaptive network. We demonstrate that a simple intuitive  rewiring rule for the
network  connections has a profound impact on the emerging network,  and is able
to generate specific network properties such as a wide degree distribution,
assortative degree correlations and the formation of two loosely connected
sub-compartments. The dynamical consequences are the emergence of new
epidemic thresholds (corresponding to first order transitions), the coexistence
of multiple stable equilibria (leading to hysteresis), and the appearance of an
oscillatory regime, all of which are absent on static SIS networks. 

We consider a network with a constant number of nodes, $N$, and 
bidirectional links, $K$. The nodes represent individuals, which are either
susceptible (S) or infected (I). In every time step and for every link
connecting an infected with a susceptible (SI-link), the  susceptible becomes
infected with the fixed probability $p$. Infected recover from the disease with
probability $r$, becoming susceptible again. In addition,  we allow susceptible
individuals to protect themselves by rewiring their links. With probability
$w$ for every SI-link, the  susceptible  breaks the link to the infected and 
forms a new link to another  randomly selected susceptible. Double- and
self-connections are not  allowed to form in this way. 

\begin{figure} 
\includegraphics[width=8cm,height=6cm]{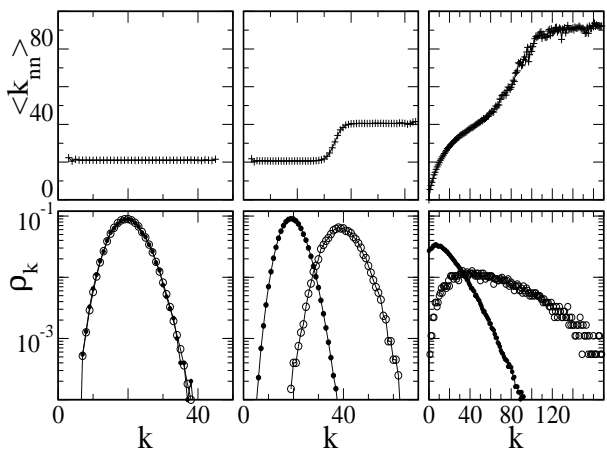}
\caption{Structure of adaptive networks. Plotted is the mean nearest-neighbor
degree $\langle k_{nn}\rangle$ (top) and the degree distribution $\rho_k$ for
susceptibles (bottom, circles) and infected (bottom, dots) depending on the
degree $k$.  (Left) Indiscriminate rewiring: the network is a random graph with 
Poissonian degree distributions and vanishing degree correlation. (Center) No
local dynamics ($p=r=0$): the infected and susceptibles separate into two
unconnected random sub-graphs. (Right) Adaptive network with rewiring and local
dynamics ($w=0.3$, $r=0.002$, $p=0.008$):  the degree distributions are
broadened considerably and a strong assortative degree correlation appears. The
plots correspond to $N=10^5$, $K=10^6$.
\label{figDegDist}}     
\end{figure}


To study the effect of adaptive rewiring consider the threshold infection probability $p^*$ that 
is necessary to maintain a stable epidemic. On a random graph without rewiring ($w=0$) the basic 
reproductive number, which denotes the secondary infections caused by a single infected node 
on an otherwise susceptible network is $R_0=p\langle k \rangle/r$, where $\langle k \rangle = 2K/N$ 
is the mean degree of the nodes \cite{Formula}. Demanding that exactly one secondary infection is 
caused yields $p^*=r/\langle k \rangle$. If rewiring is taken into account a single infected node will 
in average loose a constant fraction $w$ of its links. Therefore the degree of such a node can be written as $k(t)=\langle k \rangle \exp{(-wt)}$, where $t$ is the time since infection. By averaging over the  typical lifetime $1/r$ of an infected node, we obtain the 
threshold infection rate
\begin{equation} 
\label{eqEpiThres}
p^*=\frac{w}{\langle k \rangle (1-\exp{(-w/r)})}. 
\end{equation} 
Note, that this corresponds to $p^*=r/\langle k \rangle$ for $w=0$, but 
$p^*=w/\langle k \rangle$ for $w \gg r$. Thus, a high rewiring rate can
significantly increase the epidemic threshold and thereby reduce the prevalence
of the epidemics. 

In comparison, the effect of adaptive rewiring on the topology is more subtle.
Let us first consider the trivial case in which  rewiring is independent of the
state of the nodes (Fig.~\ref{figDegDist}, left).  In this case the degree
distribution becomes Poissonian and the average degree $\langle k_{\rm
nn}\rangle$ of the next neighbors of a given node is independent  of the degree
$k$, as one would expect in a static random graph. 

Now, assume that the adaptive rewiring  rule described above is used, but the
local dynamics is switched  off, $r=p=0$, (Fig.~\ref{figDegDist}, center). In
this case the density of infected, $i$, and susceptibles, $s=1-i$, stays
constant\cite{Notation_remark}. However, the number of SI-links is reduced systematically over time
until the network has split into  two disconnected clusters, one of which is
occupied by infected  while the other is occupied by susceptibles. Assuming that
we start with a random graph, the per-capita densities of SS-, II- and SI-links
are initially  $l_{\rm SS}=s^2\langle k \rangle/2$, $l_{\rm II}=i^2 \langle k
\rangle/2$  and $l_{\rm SI}=\langle k \rangle/2 -l_{\rm SS}-l_{\rm II}=si
\langle k \rangle$, respectively. With adaptive rewiring, in the stationary
state all SI links  have been converted into SS links so that $l_{\rm
SS}=(1-i^2) \langle k \rangle /2$ and $l_{\rm SI}=0$. Consequently, susceptibles
and infected assume different degree distributions $\rho_k$,  in which the mean
degree of a susceptible node is $\langle k_{\rm S} \rangle=(1+i) \langle k
\rangle$ and the mean degree of an infected node is $\langle k_{\rm I} \rangle=i
\langle k \rangle$. While both clusters are still individually Poissonian, the
susceptible cluster  has a higher connectivity. Since $\langle k_{\rm
nn}\rangle$ is independent of $k$ in each of the two clusters, the degree
correlation within the cluster vanishes. However a considerable net degree
correlation $r_{\rm corr}>0$ \cite{r_corr} can arise if both clusters are
considered together because $\langle k_{nn}\rangle$  is larger for the
susceptible cluster.

\begin{figure} 
\includegraphics[width=8cm,height=6cm]{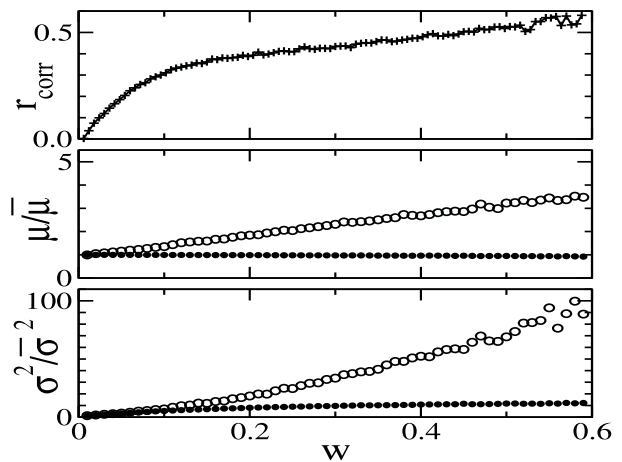}
\caption{Degree correlation index $r_{\rm corr}$ \cite{r_corr} as a function of the rewiring rate (top). Furthermore, the mean
$\mu$ (center) and the variance $\sigma^2$ (bottom) of the degree
distributions for susceptibles (circles) and the infected (dots) are shown.
Both quantities have been normalized with respect to their values in a
random graph without rewiring, $\bar{\mu}$ and $\bar{\sigma}^2$, respectively.
The plots correspond to $N=10^5$, $K=10^6$, $r=0.002$, $p=0.008$. 
\label{figNewman}} 
\end{figure}

Finally, consider the case with both adaptive rewiring and epidemic dynamics
(Fig.~\ref{figDegDist}, right). Even though rewiring is not fast enough  to
separate infected and susceptibles completely, it still structures the system
into two loosely connected clusters of susceptibles and infected (e.g., $l_{\rm
SI}\approx 0.01 \langle k \rangle$ in the figure).  While inter-cluster
connections are continuously removed by rewiring,  new ones are formed by
recoveries in the infected cluster and infections  in the susceptible cluster. 
This leads to large temporal fluctuations in the degree of a
node. As long as an individual is susceptible, its degree is increasing
approximately linear in time,  $\dot{k}=w l_{SI}$,  due to the rewiring activity
of the other susceptibles. In contrast, the degree of infected decays
exponentially,  $\dot{k}\sim - w k$.  In this way, a complicated dynamical
equilibrium can form in which the average number of inter- and intra-cluster links as well as the density of susceptibles and infected stays constant.  In this equilibrium  the continuous rewiring of connections leads to  broadened degree distributions for both infected and susceptibles  and a positive (assortative) degree correlation.

The effect of adaptive rewiring on the emerging network structure is further
quantified in Fig.~\ref{figNewman}.  With increasing $w$ the degree correlation
grows rapidly. Moreover, the mean degree of the susceptibles increases while 
the degree of the infected decreases slightly. Even more pronounced is the 
increase in the variance of the degree distribution of susceptibles, e.g.,  for
$w=0.6$ the variance $\sigma^2$ rises by a factor of $100$ in Fig.~\ref{figNewman}. This indicates the formation of strongly connected hubs and temporarily isolated nodes, which are rapidly reconnected because of rewiring. 

As we have shown adaptive rewiring promotes the isolation of infected
individuals, which can  significantly increase the epidemic threshold. However,
in doing so rewiring introduces a mixing of connections in the population and
also leads to the formation of a highly connected susceptible cluster, which is
characterised by a large variance of the degree distribution  and hence has a
lower epidemic threshold. Therefore the local effect of rewiring tends to
suppress the epidemic while the topological effect promotes it. In order to
investigate the dynamics caused by the opposing effects of rewiring  it is
useful to consider a low dimensional model. From the discussion above both the
dynamic state and the topological structure of the network can be described in
terms of the mean field  quantities $i$ and $l_{\rm SS}$ and $l_{\rm II}$. To
describe the time evolution of these variables we apply the moment closure
approximation proposed by \cite{Keeling:Moments1}. In this pair approximation
the density of all triples $l_{abc}$ in the network with the respective states
$a,b,c \in [S,I]$ are approximated by $l_{abc} =l_{ab}l_{bc}/b$,
i.e., as the product of the number of
ab-links $l_{ab}$ and the probability $l_{bc}/b$ that a given node of type $b$ has a  bc-link.  This leads to a system of three coupled ordinary differential equations 
\begin{eqnarray}
\frac{\rm d}{\rm dt}i&=& p l_{\rm SI} - r i \label{eqMoments1}\\ 
\frac{\rm d}{\rm dt}l_{\rm II}&=& p l_{\rm SI}\left(\frac{l_{\rm SI}}{s}+1\right)-2rl_{\rm II} \label{eqMoments2}\\
\frac{\rm d}{\rm dt}l_{\rm SS}&=& (r+w) l_{\rm SI} - \frac{2 p l_{\rm SI}l_{\rm SS}}{s}. \label{eqMoments3}
\end{eqnarray}
The first term in Eq.~(\ref{eqMoments1}) describes the infection of susceptible
individuals, while the second term describes recovery. These two processes also
effect the dynamics of the links. The first term in Eq.~(\ref{eqMoments2})
corresponds to the conversion of SI links into II links as a result of new
infections while the second term represents the conversion of II links into SI
links as a result of recovery. Equation~(\ref{eqMoments3}) is analogous  except
for the fact that the conversion of SI links into SS links by rewiring  has been
taken into account.

\begin{figure}
\includegraphics[width=8cm,height=9cm]{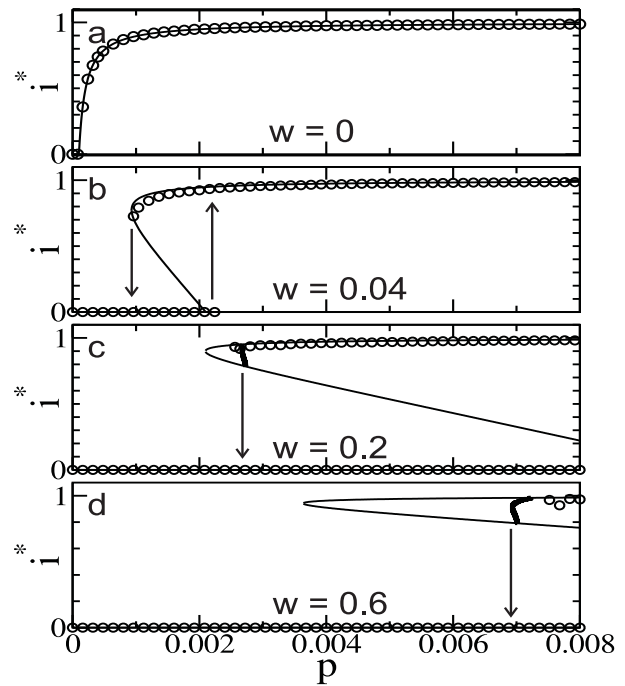}
\caption{Bifurcation diagram of the density of the infected $i^*$ as a function
of the infection probability $p$ for different values of the rewiring rate $w$.
In each diagram $i^*$ has been computed analytically from 
Eqs.~(\ref{eqMoments1}-\ref{eqMoments3}) (thin lines).  Along the stable
branches these results have been confirmed by the explicit simulation of the
full network (circles). Without rewiring only a single continuous transition
occurs at $p^*=0.0001$ (a). By contrast, with rewiring  a number of
discontinuous transitions, bistability and hysteresis loops (indicated by
arrows) are observed (b-d). Fast rewiring (c, d) leads to the emergence of
limit cycles (thick lines indicate the lower turning point of the cycles),
which have been computed numerically with the bifurcation software
AUTO\cite{AUTO}. Parameter values $N=10^5$, $K=10^6$  and $r=0.002$. 
\label{figBranches1}} \end{figure} 

In Fig.~\ref{figBranches1} the analytical results from the low order model are
compared with direct numerical simulations of the full model. Without rewiring,
there is only a single, continuous dynamical transition, which  occurs at the
well known epidemic threshold, $p^*$. As the rewiring is  switched on, this
threshold increases in perfect agreement
with Eq.~(\ref{eqEpiThres}). While the
epidemic threshold still marks the critical parameter value for invasion of new
diseases another, lower threshold, corresponding to a saddle-node bifurcation,
appears. Above this threshold an already established epidemics can persist
(endemic state). In contrast to the case without rewiring  the two thresholds
correspond to discontinuous (1st order) transitions.  Between them a region of
bistability is located, in which the healthy and  endemic state are both stable.
Thus, a hysteresis loop is formed. 

\begin{figure} 
\includegraphics[width=8cm]{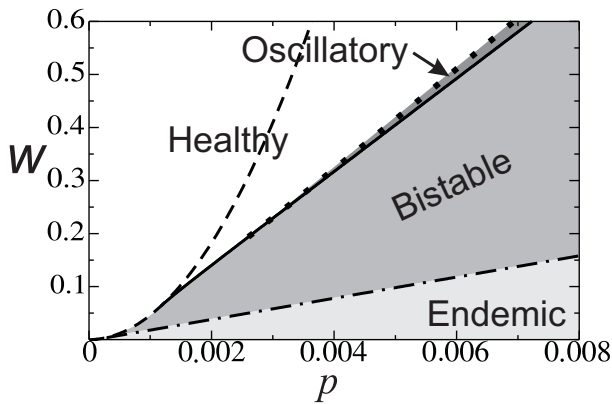}
\caption{Two parameter bifurcation diagram showing the dependence on the
rewiring rate $w$ and the infection probability $p$. In the white and light grey
regions  there is only a single attractor, which is a healthy state in the
white  region and an endemic state in the light grey region. In the medium grey
region  both of these states are stable. Another smaller region of bistability
is shown  in dark grey. Here, a stable healthy state coexists with a stable
epidemic cycle.  The transition lines between these regions correspond to
transcritical (dash-dotted),  saddle-node (dashed), Hopf (continuous) and cycle
fold (dotted) bifurcations. The transcritical bifurcation line agrees very well 
with Eq. (1). Note, that the saddle-node and transcritical
bifurcation lines emerge from a cusp bifurcation at $p=0.0001$,$w=0$.
Parameters as in Fig.\ref{figBranches1}.
\label{figBranches2}} \end{figure}

Our numerical simulations show that the presence of a hysteresis loop and first
order transitions is a generic feature of the adaptive model and can be observed
at all finite rewiring rates. While increasing the rewiring rate hardly
reduces the  size of the epidemic in the endemic state, the nature of the
persistence threshold changes  at higher rewiring rates. First, a subcritical
Hopf bifurcation, which gives rise to an  unstable limit cycle replaces the
saddle-node bifurcation. At even higher  rewiring rates this Hopf bifurcation
becomes supercritical. Since the emerging  limit cycle is now stable, the Hopf
bifurcation marks a third  threshold at which a continuous transition to
oscillatory dynamics occurs. However, these oscillations can only be observed in
a relatively small parameter region (s.~Fig.~\ref{figBranches2}) before the
persistence threshold is encountered,  which now corresponds to a fold
bifurcation of cycles.

In summary, we have shown that the interplay between the dynamics and topology
of an adaptive network can give rise to rich dynamics.  Here we have studied
only  the simplest example of an adaptive network, in which the number of nodes
and links remains  constant and the local dynamics is simple. Nevertheless, the
adaptive nature of the system gives rise to dynamical features, like 
bistability and cycles. While we observe stable oscillations only in a small
parameter region,  the fact that they already appear in this simple example
indicate that they can also be expected in more complex models. In fact, we have
found much larger oscillatory  regions in other model variants with different
rewiring rules. Besides epidemic dynamics our findings also have strong
implications for the spreading of information, opinions  and beliefs in a
population, which can be described in a similar way. 

For the control of real world diseases adaptive rewiring is beneficial since  
it increases the invasion threshold and also the persistence threshold for
epidemics.  However, the topological changes that are induced as a natural
response to an emerging  disease are cause for concern. The topology of at the
peak of a major epidemic can be very different from that in the disease-free
state. In particular, positive degree correlations  can rapidly arise, reducing
effectiveness of targeted vaccination. Further, the formation  of a densely
linked cluster of susceptibles at high infection densities can enable the 
persistence of diseases which would not be able to persist at low infection
densities.   Therefore, a disease which seems to be a minor problem while it is
rare can be very  difficult to combat once it has reached an endemic state.


\begin{acknowledgments} 
This work was supported by the Volkswagen-Stiftung.
\end{acknowledgments} 


\begin{thebibliography}{16}
\bibitem{Notation_remark}
%
%
We use upper (lower) case letters to denote the state 
(density) of a node.
\bibitem{r_corr}
In undirected networks the degree correlation can be computed as
$r_{\rm corr}=\sigma_q^{-2}\sum_{ab} ab (e_{ab}-q_a q_b)$ where 
$q_a=(a+1)\rho_{a+1}/\sum_k k \rho_k$,
$e_{ab}$ is the probability that a randomly chosen link connects
nodes with the degrees $a+1$ and $b+1$, and $\sigma_q^2$
is the variance of the distribution $q_a$ \cite{Newman:Assortative}.
\end{thebibliography}

\end{document}